 \documentclass{emulateapj}

\usepackage{color}

\newcommand{\be}{\begin{equation}}
\newcommand{\ee}{\end{equation}}
\newcommand{\ber}{\begin{eqnarray}}
\newcommand{\eer}{\end{eqnarray}}
\newcommand{\bers}{\begin{eqnarray*}}
\newcommand{\eers}{\end{eqnarray*}}

\shorttitle{Turbulent Damping of Alfven waves}
\shortauthors{Lazarian}

\begin{document}

\title{Damping of Alfven waves by Turbulence and its Consequences: from Cosmic-Rays Streaming to Launching Winds}

\author{A. Lazarian}
\affil{Department of Astronomy, University of Wisconsin-Madison, 475 Charter St., Madison, WI 53705}

\begin{abstract}
This paper considers turbulent damping of Alfven waves in magnetized plasmas. We identify two cases of 
damping, one related to damping of cosmic rays streaming instability,
the other related to damping of Alfven waves emitted by a macroscopic wave source, e.g. stellar atmosphere. 
The physical difference between the two cases is that in the former case the generated waves are emitted in respect to the local direction of magnetic field, in the latter in respect to the mean field. The scaling of damping is
different in the two cases. We the regimes of turbulence ranging from subAlfvenic to superAlfvenic we obtain analytical expressions for the damping rates and define the ranges of applicability of these expressions. Describing the damping
of the streaming instability, we find that for subAlfvenic turbulence the range of cosmic ray energies influenced by weak turbulence is unproportionally large compared to the range of scales that the weak turbulence is present. On the contrary, the range of cosmic ray energies affected by strong Alfvenic turbulence is rather limited. A number of astrophysical applications of the process ranging from launching of 
stellar and galactic winds to propagation of cosmic rays in galaxies and clusters of galaxies is considered.
 In particular, we discuss how to reconcile the process of turbulent damping with the observed isotropy of the 
Milky Way cosmic rays.

\end{abstract}


\section{Introduction}

Astrophysical plasmas are magnetized and turbulent (see McKee \& Ostriker 2007, Sharma et al. 2009, Brandenburg \& Lazarian 2013). The propagation of Alfvenic waves in MHD turbulence is a problem of great astrophysical significance with applications to key astrophysical processes (see Uhlig et al. 2012, Wiener, Oh \& Guo 2013, van der Holst et al. 2014, Lynch et al. 2014).  Waves in astrophysical environments can arise from instabilities, e.g. from the cosmic ray (CR) instability due to streaming (Lerche 1967, Kulsrud \& Pearce 1969, Wentzel 1969, Skilling 1971). They can also be induced in the environment by macroscopic sources, e.g. they can be the result of vibrations of the stellar surface or arise from magnetic reconnection (see Konigl 2009 and ref. therein, Suzuki 2013).  The most well-known consequences of wave propagation and damping range from heating of coronal gas in Solar atmosphere (e.g. Arber, Brady \& Shelyag 2016,  Reep \& Russell 2016), acceleration and scattering of cosmic rays (e.g. Jokipii 1966, Schlickeiser 2002, 2003), and launching of solar and stellar winds (e.g. Suzuki \& Inutsuka  2005, van Ballegooijen, \& Asgari-Targhi 2016). The damping of waves induced by CR streaming is important for the confinement and acceleration of CRs  in interstellar, intergalactic and interplanetary environments (see Ensslin et al. 2011,  Badruddin, \& Kumar, A. 2016, Kulsrud 2005). In more general terms, the damping must be understoond in order to answer the long-standing question of the relative importance of waves and turbulence in various astrophysical settings (see Petrosian 2015).  

Initial studies of Alfven wave damping in turbulent plasmas were done in Silimon \& Sudan (1989) in a model of isotropic turbulence which does not correspond to the present understanding of magnetized turbulence (see \S 2 for more discussion of MHD turbulence). More recently, turbulent damping of Alfven waves was suggested as a process for suppressing of CR streaming instability in Yan \& Lazarian (2002, henceforth YL02). This process was quantified in an important study by Farmer \& Goldreich (2004, henceforth FG04) for the original Goldreich \& Sridhar (1995, henceforth GS95) model of Alfvenic turbulence and was tested numerically in Beresnyak \& Lazarian (2008). 

FG04 employs the original  GS95 model of {\it strong} Alfvenic turbulence that assumes that turbulence is injected {\it isotropically} with velocity $u_L$ that is equal to the Alfven velocity $V_A$. Although important, this regime does not cover the full variety of astrophysical conditions. Moreover, the work deals only with the Alfven waves generated by CR streaming instability and does not cover Alfven waves from the external macroscopic source. At the same time,
the propagation of Alfven waves in turbulent plasmas is important for understanding of heating of stellar atmospheres, galactic halos, intracluster media (see Zhuravleva et al. 2014). The momentum deposited by Alfven waves in plasmas is important for 
launching stellar and galactic winds. 

 First of all, as we discuss further in the paper, the damping of Alfven waves generated by streaming instability is different from the damping of  Alfven wave generated by a microscopic source. Moreover, strong isotropically injected 
 MHD turbulence discussed in FG04 is one of
 the types of MHD turbulence that is relevant to astrophysical settings. Depending on the injection scale, injection velocity  and media magnetization, magnetized turbulent motions can demonstrate regimes of isotropic superAlfvenic turbulence, extremely anisotropic weak turbulence, and strong subAlfvenic turbulence with anisotropic injection. Turbulent motions at the injection scale can
 also damp Alfven waves.    
 
  The quantitative studies of the damping of the Alfven waves injected with respect to (a) local magnetic fields and (b)  global mean field by MHD turbulence in the aforementioned different regimes, as well as with the turbulence at scales larger than the turbulence injection scale, is the goal of this paper. 

In particular, in the present paper we address the damping of Alfven waves by turbulence taking into account both the spatial structure of the front of the Alfven wave and what are the actual properties of MHD turbulence with which these waves are interacting. The former depends on how the Alfven waves were generated, the latter depends on the properties of MHD turbulence in different regimes. As we mentioned earlier, the FG04 study addresses the damping of Alfven waves generated by the streaming instability of cosmic rays (see Kulsrud 2005, Longair 2011), which corresponds to the waves generated with respect to the local direction of the magnetic field that is sampled by the streaming particles. The distinction between the local and global systems of reference (Lazarian \& Vishniac, 1999, henceforth LV99, Cho \& Vishniac 2000, Maron \& Goldreich 2001, Cho, Lazarian \& Vishniac 2002) plays a crucial role in the present day theory of MHD turbulence and, as we will show later, the damping of Alfven waves differs in the case of Alfven waves launched with respect to the local magnetic field and with respect to the mean magnetic field. This important distinction has not been addressed in the earlier studies. Similarly, we are not aware of quantitative studies of Alfven wave damping by MHD turbulence in different regimes. 

We believe that providing the study of the Alfven wave damping for a wide range of possible turbulent astrophysical settings is very important. Indeed, magnetic turbulence in stellar and pulsar atmospheres, in halos of spiral galaxies MHD is subAlfvenic with the Alfven Mach number $M_A=u_L/V_A<1$. In spiral arms, different parts of the interstellar media exhibit different degrees of magnetization (Draine 2011). For instance, molecular clouds may be superAlfvenic (see Luntilla et al. 2008), although this is not a universally accepted opinion (see Li \& Henning 2011). Turbulence in clusters of galaxies is generally accepted to be superAlfvenic (see Brunetti \& Lazarian 2007, Brunetti \& Jones 2014,  Brunetti 2016).   

In what follows, in \S 2 we derive the relations for different regimes of the MHD turbulence that we employ in this paper. By providing this derivation we clarify the points that are essential for our further explanation of the nature of Alfven wave damping. Then we provide a physical picture of damping in \S 3. Historically, the damping of streaming instability by turbulence was the first to be addressed (YL02, FG04). This type of damping is associated with Alfven waves generated with respect to the local magnetic field. Therefore, in \S 4 we quantify the damping of waves generated in the local system of reference for both subAlfvenic and superAlfvenic turbulence. There we also define the ranges of wavelength for which the damping by weak and strong subAlfvenic turbulence are applicable as well as define the range of applicability of damping by superAlfvenic turbulence.  We later consider the second case of damping of Alfven waves, i.e. the case of damping of waves generated by a macroscopic source. Therefore, for the aforementioned variety of MHD turbulence regimes, we consider the case of damping of Alfven waves that are injected with respect to the mean field in \S 5. We compare the turbulent damping of Alfven waves with the non-linear Landau damping of Alfven waves in \S 6 and briefly outline some of the astrophysical implications of our study in \S7. We compare our results with those
in earlier works in \S 8. Our discussion is provided in \S 9 and our summary is given in \S  10.

\section{Regimes of Alfvenic turbulence}

MHD theory is applicable to astrophysical plasmas at sufficiently large scales
and for many astrophysical situations the requirement for Alfvenic turbulence to be applicable is that the turbulence is studied at scales
substantially larger than the ion gyroradius $\rho_i$ (see more in Kulsrud 2005, Eyink et al. 2011).  
In what follows we consider this criterium satisfied. For Alfvenic  turbulence velocity and magnetic field fluctuations have the same scaling and thus below we focus on velocity scaling. 

MHD turbulence can be decomposed into the cascades of Alfven, slow and fast modes (GS95, Lithwick \& Goldreich 2001) the concept that has been elaborated and proved numerically (Cho \& Lazarian 2002, 2002, Kowal \& Lazarian 2010, Takamoto \& Lazarian 2016). For non-relativistic turbulence the Alfvenic cascade is marginally affected by two other fundamental MHD modes (see Cho \& Lazarian 2002, Takamoto \& Lazarian 2016, cf. Stone et al. 1998) and therefore we focus on Alfvenic turbulence that is responsible for Alfven wave non-linear damping. 

The the pioneering studies of Alfvenic turbulence were done by Iroshnikov (1964) and Kraichnan (1965) for a hypothetical model of isotropic MHD turbulence.  Later studies (see Montgomery \& Turner 1981, Matthaeus et al,1983, Shebalin et al 1983, Higdon 1984) discovered the anisotropic nature of the energy cascade and paved the way for further
advancements in the field. The breakthrough work by GS95 provided the theory of trans-Alfvenic turbulence, i.e. turbulence corresponding to injection velocity $u_L$ at the injection scale $L$ equal to Alfven velocity $V_A$ which corresponds to the Alfven Mach number
\be
M_A=\frac{u_L}{V_A},
\label{Alfven_mach}
\ee
equal to unity. The generalization of the GS95 theory for $M_A<1$ and $M_A>1$ was obtained later (LV99, Lazarian 2006). Note that the original GS95 theory was also augmented by the concept of local systems of reference (LV99, Cho \& Vishniac 2000, Maron \& Goldreich 2001, Cho, Lazarian \& Vishniac 2002)
that specifies that the turbulent motions should be viewed not in the system of reference of the mean magnetic field, but in the local system of reference of the turbulent eddies. This is quite a natural concept, which, however, was missed by the earlier studies. Indeed, for the small scale turbulent motions the only magnetic field that matters is the magnetic field in their vicinity. Thus this local field, rather than the mean field, should be considered. Keeping this in mind, in what follows, we describe turbulent motions using not parallel and perpendicular wave numbers, but parallel to local magnetic field size of the eddy $l_{\|}$ and perpendicular scale $l_{\bot}$. The distinction between the local system of reference related to the local field and the global system of reference related to the mean field is also important for the two cases of Alfven wave damping that we consider in this work.

 In Table~1 (see  \S 9) we present various regimes of Alfvenic turbulence and the ranges for which those are applicable. 
 Below we consider first subAlfvenic turbulence corresponding corresponding the Alfven Mach number $M_A<1$.
 For Alfvenic perturbations the relative perturbations of velocities and magnetic fields are related in the
following way:
\be
\frac{\delta B_l}{B}=\frac{\delta B_l}{B_L}\frac{B_L}{B}=\frac{u_l}{u_L} M_A=\frac{u_l}{V_A},
\label{same}
\ee
where $B_l$ is the perturbation of the magnetic field $B$ at the scale $l$, $B_L$ is the perturbation of the magnetic field at the injection scale $L$, while $u_l$ is the velocity fluctuation at the scale $l$ in the turbulent flow with energy injected with the velocity $u_L$.

To understand the nature of non-linear damping of Alfven waves, it is useful to get an insight into a detailed picture of
cascading by Alfven turbulence (see Cho, Lazarian \& Vishniac 2003 for more details). Consider colliding Alfvenic wave packets with parallel scales $l_{\l}$ and perpendicular scales $l_{\bot}$. The change
of energy per collision is
\begin{equation}
 \Delta E \sim (du^2_l/dt) \Delta t,
 \label{init}
 \ee
 where the first term represents the change of the energy of a packet in time as it interacts with 
 the oppositely moving wave packet. Naturally, the time of the interaction is the time of the passage of
 the given wave packet through the oppositely moving wave packet of the size $l_{\|}$, thus the interaction time
 $\Delta t \sim l_{\|}/V_A$. To estimate the characteristic rate of cascading one should accept that the cascading 
 of our wave packet is happening due to the change of the structure of the oppositely moving wave packet which
 is happening with the rate $u_l/l_{\bot}$. Indeed, more structures are being created in the passing package as the
 background field affected by the opposite package evolves. Therefore Eq. (\ref{init}) becomes
 \be
  \Delta E 
 \sim {\bf u}_l \cdot \dot{\bf u}_l\Delta t
 \sim  (u_l^3/l_{\perp}) (l_{\|}/V_A),
 \label{change}
\end{equation}

The fractional energy change per collision is the ratio of $\Delta E$ to $E$,
\be
  f \equiv \frac{\Delta E}{u^2_l}
                           \sim \frac{ u_l l_{\|} }{ V_A l_{\perp} },
                         \label{fraction}
\ee
which is the measure of  the strength of the nonlinear interaction. $f$ is the ratio of the rate of shearing of the wave packet
$u_l/l_{\bot}$ to the rate of the propagation of the wave packet $V_A/l_{\|}$. If the shearing rate is much smaller than the propagation rate, for $f\ll1$ 
the cascading is
a random walk process which means that
\be
\aleph=f^{-2},
\label{aleph}
\ee
steps are required for the cascading, i.e. the cascading time is 
\begin{equation}
t_{cas}\sim \aleph \Delta t .
\label{tcas}
\end{equation}
For $\aleph>1$, the turbulence is  called {\it weak} and if $\aleph\approx 1$ the turbulence is called strong. 

The Alfvenic 3-wave resonant interactions are governed by relations for wavevectors
that reflect momentum conservation
 $ {\bf k}_1 + {\bf k}_2  = {\bf k}_3$,  and 
 frequencies reflecting energy conservation $\omega_1 +  \omega_2  = \omega_3$.
 With Alfven wave packets interacting with oppositely moving ${\bf k}$ and have the 
 dispersion relation $\omega = V_A |k_{\|}|$, 
 where $k_{\|}\sim l_{\|}^{-1}$ is the component of wavevector
parallel to the background magnetic field the increase of $k_{\bot}\sim l_{\bot}^{-1}$ occurs.  The decrease of
$l_{\bot}$ with $l_{\|}$ being fixed  signifies the increase of the energy change per collision. This eventually makes
$\aleph$ of the order of unity and the approximation of the weak Alfvenic turbulence breaks down. For $\aleph\approx 1$
the GS95 critical balanced condition
\be
u_l l_{\bot}^{-1}\approx V_A l_{\|}^{-1},
\label{crit}
\ee
is satisfied with the cascading equal to the wave period $\sim \Delta t$. The value of $\aleph$ cannot
decrease further and the turbulence evolves as  {\it strong Alfvenic turbulence}. Therefore the further decrease of $l_{\bot}$ entails
the corresponding decrease of $l_{\|}$ to keep the critical balance satisfied. The ability to change $l_{\|}$ means that 
the frequencies of interacting waves increase, which is possible as cascading introduces the uncertainty in
wave frequency $\omega$ of the order of $1/t_{cas}$. 

The cascading turbulent energy flux for incompressible fluid is (Batchelor 1953):
\begin{equation}
\epsilon\approx u_l^2/t_{cas}=const,
\label{cascading}
\end{equation}
which in the hydrodynamic case provides
\be
\epsilon_{hydro}\approx u_l^3/l\approx u_L^3/L=const,
\ee
where the relation $t_{cas}\approx l/u_l$ is used.

For the weak cascade $\aleph \gg 1$ provides (LV99)
\be
\epsilon_w \approx \frac{ u_l^4} {V_A^2 \Delta t (l_{\bot}/l_{\|})^2} \approx \frac{u_L^4}{V_A L},
\label{eps_weak}
\ee
where Eqs. (\ref{cascading}) and (\ref{tcas}) are used. The second relation in Eq. (\ref{eps_weak}) follows from the assumed isotropic injection of turbulence at the scale $L$. 

Taking into account that
$l_{\|}$ is constant, it is easy to see that Eq. (\ref{eps_weak}) provides
\be
u_l\sim u_L (l_{\bot}/L)^{1/2},
\label{u_weak}
\ee
which is different from the Kolmogorov $\sim l^{1/3}$ scaling.\footnote{Using the relation $k E(k) \sim u_k^2$ it is easy to see that the spectrum of weak turbulence is $E_{k, weak}\sim k_{\bot}^{-2}$ (LV99, Galtier et al. 2000).} 

For the accepted model of isotropically injected turbulence at scale $L$, the initial $l_{\|}=L$ and
the transition to $\aleph\approx 1$, i.e. to strong turbulence, occurs (LV99)
\begin{equation}
l_{trans}\sim L(u_L/V_A)^2\equiv L M_A^2.
\label{trans}
\end{equation} 
Thus, weak turbulence has a limited,  i.e. $[L, L M_A^2]$, inertial range and at scales less than $LM_A^2$
it transits into the regime of strong turbulence. The velocity corresponding to the transition follows
from $\aleph\approx 1$ condition given by Eqs. (\ref{aleph}) and (\ref{fraction}): 
\be
u_{trans}\approx V_A \frac{l_{trans}}{L}\approx V_A M_A^2.
\label{vtrans}
\ee

The relations for the strong turbulence in the subAlfvenic 
regime obtained in LV99 can be easily derived as follows. Indeed, the turbulence becomes strong and
cascades over one wave period, which according to Eq. (\ref{crit}) is equal to $l_{\bot}/u_l$.  
Substituting the latter in Eq. (\ref{cascading}) one gets
 \be
 \epsilon_s\approx \frac{u_{trans}^3}{l_{trans}}\approx \frac{u_l^3}{l}=const, 
 \label{alt_sub2}
 \ee
 which is analogous to the hydrodynamic Kolmogorov cascade in
the direction perpendicular to the local direction of the magnetic field. This cascade starts at $l_{trans}$ and
has the injection velocity given by Eq. (\ref{vtrans}). Thus (LV99)
\begin{equation}
u_{l}\approx V_A \left(\frac{l_{\bot}}{L}\right)^{1/3} M_A^{4/3}.
\label{vll}
\end{equation}
In terms of the injection velocity $u_L$ Eq. (\ref{vll}) can be rewritten as
\be
\delta u_{l}\approx u_L \left(\frac{l_{\bot}}{L}\right)^{1/3} M_A^{1/3}.
\label{alternative}
\ee 
Substituting the latter expression in Eq. (\ref{crit}) one gets the relation between the parallel and perpendicular
scales of the eddies (LV99):
\begin{equation}
l_{\|}\approx L \left(\frac{l_{\bot}}{L}\right)^{2/3} M_A^{-4/3}.
\label{Lambda1}
\end{equation}
The relations given by Eq. (\ref{Lambda1}) and (\ref{vll}) reduce to the well-known GS95 scaling for transAlfvenic turbulence if 
$M_A\equiv 1$. 

The {\it superAlfvenic} turbulence corresponds to $u_L > V_A$, which is equivalent to $M_A>1$. For $M_A\gg 1$ the turbulence at scales close
to the injection scale is essentially hydrodynamic
as the influence of magnetic forces is of marginal importance, i.e. the velocity is Kolmogorov
\be
u_l=u_L (l/L)^{1/3}.
\label{u_hydro}
\ee
The cascading nature changes at the scale 
\be
l_{A}=LM_A^{-3}, 
\ee
at which the turbulent velocity becomes equal to the Alfven velocity (Lazarian 2006).  The rate of cascading for $l<l_A$ can be written as:
\be
\epsilon_{superA}\approx u_l^3/l\approx M_A^3 V_A^3/L=const. 
\label{super_alt1}
\ee
This cascading can be related to the GS95 cascade, if the scale given by Eq. (\ref{super_alt1}) is taken as
the injection scale for the transAlfvenic turbulence and the corresponding scaling follows from  Eq. (\ref{Lambda}) and (\ref{vll})  for the injection of $V_A$. In other words:
\begin{equation}
l_{\|}\approx L \left(\frac{l_{\bot}}{L}\right)^{2/3} M_A^{-4/3},
\label{Lambda}
\end{equation}
\begin{equation}
u_{l}\approx u_{L} \left(\frac{l_{\bot}}{L}\right)^{1/3} M_A^{1/3}.
\label{vl}
\end{equation}
The relations above are used in the discussion of the Alfven wave damping that follows.

\section{General remarks on turbulent damping of Alfven waves}

Linear Alfven waves propagate without inducing {\it irreversible} distortions on each other. The situation changes when Alfven waves interact with Alfvenic turbulence. There the structure of magnetic field lines changes significantly over the time of propagation of the wave and this causes the non-linear distortion. As a result, the waves undergo cascading and dissipate
together with turbulence. The difference between this process and  the other wave damping processes that turbulent damping does
not depend on plasma microphysics.  

We provide calculations further in this paper, but here we provide some simple arguments explaining the physics of the two regimes of damping that we consider further along. The dynamically important magnetic field of the turbulent fluid is aligned better locally on the scale of the small eddies. Therefore the Alfven waves emitted parallel to the local magnetic field, as this is the case of Alfven waves emitted by the streaming instability, will experience the least distortions from the oppositely moving eddies. Formally, the wave moving exactly parallel to the magnetic field corresponds to a wave packet with $l_{\bot}=\infty$. Therefore the least distorted Alfven waves can be described as the Alfven packages having the largest value of $l_{\bot}$. The wave packages are most efficiently distorted by the oppositely moving wave packages with the same $l_{\bot}$. The larger $l_{\bot}$ the longer time for the evolution of the corresponding packages, e.g. for strong GS95 turbulence it corresponds to $l_{\bot}/v_l\sim l_{\bot}^{2/3}$. We will
show below that for the given Alfven wave wavelength the largest $l_{\bot}$ corresponds exactly to the case of emission parallel to the local direction of the magnetic field. This is the case of Alfven wave damping relevant to the streaming instability. The other case is the Alfven wave emission by a macroscopic source. If Alfven waves are emitted at an angle $\theta$ to the magnetic field, the value of $l_{\bot}$ is smaller and therefore the turbulent damping of Alfven waves is faster. This is also the case of, for instance, if Alfven waves are emitted parallel to the mean field. There the dispersion of the magnetic field direction in respect to the mean field acts as the angle $\theta$. Naturally, the damping of Alfven waves depends on the regime of turbulence that the waves interact with. For instance, if the turbulent perturbations are weakly non-linear, as in the case of weak turbulence, these perturbations should induce damping which is slower compared with the strongly non-linear perturbations of the strong turbulence.

In what follows we assume that the propagating Alfven waves are weak enough so that they do not distort the Alfvenic turbulence. In the opposite case we expect the turbulence to get more features of  the turbulence with non-zero cross helicity which is frequently called "imbalanced turbulence". A tested model of such turbulence in Beresnyak \& Lazarian (2008b) predicts a significant decrease of cascading for the strong wave packages. We may expect that the perturbations significantly stronger than the background turbulence can disturb the background turbulence inducing its imbalance. Thus the waves can potentially propagate over longer distances compared with the estimates that we are going to obtain below.

\section{Damping of the streaming instability}

In this section we consider the damping of Alfvenic waves that are generated with respect to the local magnetic field direction. This sort of damping is associated with the damping of the streaming instability of cosmic rays (YL02, FG04). The damping of Alfven waves in the global system of reference relevant to the damping of Alfven waves emitted by macroscopic sources is considered in \S 5. 

\subsection{Streaming instability and local system of reference}
 
 Alfven waves can be generated by different astrophysical sources. The streaming instability of CRs is an important process that generates Alfven waves. The generation happens as particles interact with the local magnetic field and the sampling scale for the magnetic field is the Larmor radius of the energetic particle. This is a typical situation when one must consider the local system of reference related to the local direction of the wondering magnetic field (LV99, Cho \& Vishniac 2000, Maron \& Goldreich 2001, Cho, Lazarian \& Vishniac 2002).  
 
 The growth rate of the streaming instability for the direction parallel to the local direction of the magnetic field is given by the expression (see Kulsrud \& Pearce 1969):
 \be
 \Gamma_{cr} \approx \Omega_B \frac{n_{cr}(>\gamma)}{n_i}\left(\frac{v_{stream}}{V_A} -1 \right),
 \label{gamm_cr}
 \ee
 where $\Omega_B=eB/mc$ is the particle gyrofrequency, $n_{cr}$ is the number density of CRs with gyroradius $r>\lambda=\gamma mc^2/eB$. The streaming velocity $v_{stream}$ enters Eq. (\ref{gamm_cr}) as a ratio with $V_A$. 
 
 For the instability to operate, the growth rate given by Eq. (\ref{gamm_cr}) should exceed the rate of the turbulent damping that we quantify below. 

\subsection{Damping by SubAlfvenic strong turbulence}

We consider first the case of strong subAlfvenic turbulence. Treating subAlfvenic strong turbulence as a test case, we consider different ways of deriving the result. 

The first approach that we present is based on calculating the distortion of the wave by evolving turbulent fluctuations as the waves propagate along magnetic field lines. The distortion of the wavefront arises from the magnetic field lines wondering over angle $\theta_x$. This angle depends on the fluctuations of the magnetic field $\delta B_x$ induced by turbulence with perpendicular scale $x$. Simple geometric considerations suggest that the distortion induced by a wave propagating along magnetic field for the time $t$ is
\be
\delta_x\approx V_A t \sin^2\theta_x \approx V_A t \left(\frac{\delta B_x}{B} \right)_t^2, 
\label{deltax}
\ee
where the perturbation induced by turbulence evolves as
\be
\left(\frac{\delta B_x}{B}\right)_t\approx \left(\frac{u_x}{V_A}\right) \left(\frac{t}{x/u_x}\right),
\label{deltaB}
\ee
where $u_x$ is the velocity corresponding to the magnetic field fluctuation $\delta B_x$. The time $t$ in Eq. (\ref{deltaB}) is less than the eddy turnover time $x/u_x$ and the ratio reflects the partial sampling of the magnetic perturbation by the wave.
Substituting the scaling of strong subAlfvenic turbulence for $u_x$ in Eq. (\ref{deltaB}) one can rewrite Eq. (\ref{deltax}) as
\be
\delta_x\approx \frac{V_A^3 M_A^{16/3} t^3}{x^{2/3} L^{4/3}}.
\label{deltax2}
\ee

The damping of the wave with the wavelength $\lambda$ corresponds to the "resonance condition" $\delta_x=\lambda$ and substituting this in Eq. (\ref{deltax2}) one can express the
perpendicular scale of the "resonance" magnetic perturbations that distort the wave:
\be
x\approx \frac{V_A^{9/2} t^{9/2} M_A^8}{\lambda^{3/2} L^2}.
\label{x1}
\ee
The required time for the damping is equal to the turnover of the resonant eddy:
\be
t\approx\frac{x}{u_l}\approx\frac{V_A^2 t^3 M_A^4}{\lambda L},
\ee
which gives the rate of turbulent damping of Alfven waves
\be  
\Gamma_{subA} \approx t^{-1},
\label{t}
\ee
or
\be
\Gamma_{subA, s} \approx \frac{V_A M_A^2}{\lambda^{1/2} L^{1/2}}.
\label{gamma1}
\ee
For transAlfvenic turbulence, i.e. $M_A=1$ this result transfers to the one in FG04. We point out, however, the square of the Alfven Mach number dependence, which means a {\it significant} change of the damping rate for subAlfvenic turbulence. We also note that in FG04 the injection scale for turbulence was defined not as the actual injection scale, but the scale at which the turbulent velocity becomes equal to the Alfven one. We discuss the implications of this in
\S 8 where we compare our approach/results with those in FG04. 

For isotropic injection of turbulence the maximal perpendicular scale of strong subAlfvenic motions is given by $x_{max}=LM_A^2$. Therefore, if one substitute this in Eq. (\ref{x1}) and simultaneously uses Eq. (\ref{t}) and Eq. (\ref{gamma1}) to express $t$, one gets
\be
\lambda_{max,s}\approx LM_A^4.
\label{max1}
\ee
For the streaming instability the particles emit Alfven waves of the order of the particle gyroradius $r_L$. Therefore the range of
$r_L$ is limited to
\be
r_L<LM_A^4,
\ee
which can be a serious limitation if $M_A$ is sufficiently small. The larger energy particles the interactions happen with weak
turbulence. We discuss this regime of damping in \S 4.3, while below we provide another derivation of Eq. (\ref{gamma1}).

Because of the significance of wave damping it is also useful to present a more intuitive derivation of the same result that is based on the notion of propagating wave packets that we employed obtaining Eq. (\ref{change}).  Consider two oppositely moving packets with the perpendicular scale $x'\sim k_{\bot}'^{-1}$. Each packet induces magnetic field distortion $\theta_x'$ of the oppositely moving waves. Consider a locally emitted Alfven wave moving parallel to the local direction of magnetic field with wavenumber $k_{\|}^{-1}\sim \lambda$. It is easy to see that  the wave gets distorted by interacting with turbulence with 
the perpendicular  $k_{\bot} \sim k_{\|}\sin \theta_x'$. The interactions of a wave with $k_{\bot}$ and the oppositely moving packages will be most efficient if it is "resonant" i.e. $k_{\bot}'=k_{\bot}$.\footnote{It is possible to show that the interactions with smaller and larger turbulent scales is subdominant compared with the interaction with the "resonant" scale.} This suggests the relation $k_{\|}\sin \theta_x=k_{\bot}$, which determines the perpendicular scale of the wave package which will cascade the wave
\be
\lambda \approx x \sin \theta_x\approx x \frac{\delta B_x}{B}.
\label{alt_lambda}
\ee
Inserting the scaling given by Eqs. (\ref{deltaB}) and (\ref{vll}) it is possible to get the expression for the "resonant" perpendicular scale $x$:
\be
x=L^{1/4} \lambda^{3/4} M_A^{-1},
\ee
which can then be used to find the rate of damping defined as $\Gamma_{subA, s}\approx u_x/x$, which reproduces the earlier result given by Eq. (\ref{gamma1}). Within this approach the maximal wavelength of the Alfvenic wave that can be damped by strong subAlfvenic turbulence can be obtained from Eq. (\ref{alt_lambda}) if the scale $l_{trans}$ is used instead of $x$, i.e.
\be
\lambda_{max,s}\approx \left(\frac{u_{trans}}{V_A}\right) l_{trans}\approx LM_A^4,
\label{max_s}
\ee
which coinsides with the result given by Eq. (\ref{max1}). The minimal scale of waves that are being damped depend on
the perpendicular scale of the smallest Alfvenic eddies $l_{min}$. Using Eq. (\ref{alt_lambda}) and the scaling of strong turbulence
given by Eq. (\ref{vll}) one can get the range of $r_L$ affected by turbulent damping arising from strong subAlfvenic turbulence:
\be
\frac{l_{min}^{4/3}}{L^{1/3}}M_A^{4/3}<r_L<LM_A^4,
\label{min_max_s}
\ee
which indicates that the waves much smaller than $l_{min}$ can be damped. The value of $l_{min}$ can be large in
partially ionized gas (see Xu et al. 2015). Due to the differences of $r_L$ for protons and electrons Eq. (\ref{min_max_s}) presents
a situation when the streaming instability of electrons is damped, while it is damped for protons. 

The damping of streaming instability for $r_L< \frac{l_{min}^{4/3}}{L^{1/3}}M_A^{4/3}$ is present, but significantly reduced. 
An estimate of it can be obtained by considering the distortion $\delta_x\ll \lambda$ given by Eq. (\ref{deltax2}) 
for the time period of
the wave $\lambda/V_A$, which is significantly less than the period of the eddy at the scale $l_{min}$, $t_{eddy}\approx 
l_{min}^{2/3} L^{1/3}/(V_A M_A^{4/3})$. The distortions accumulate as a random walk with the time step given by $t_{eddy}$.
The damping requires $\lambda/\delta_x$ steps, which results in the damping rate
\be
\Gamma_{sub, s, r_L\ll l_{min}}\approx \frac{M_A^{12} V_A r_L^4}{l_{min}^2 L^3},
\ee
which also illustrates inefficiency of damping by turbulence with the perpendicular scale larger than the "resonant" scale.

\subsection{Damping by subAlfvenic weak turbulence}

For waves longer than $\lambda_{max, s}$ the wave is cascaded through weak interactions together with the corresponding wavepackets, the perpendicular
wave scales for which are given by Eq. (\ref{alt_lambda}). The difference here, however, is that the scaling of weak turbulence given by Eq. (\ref{u_weak}) should be used. This gives the relation between the Alfven wave wavelength and the perpendicular scale of the "resonant" weak mode $l_{\bot}$
\be
\lambda=l_\bot \left(\frac{l_{\bot}}{L}\right)^{1/2}M_A,
\label{lambda_w}
\ee
which provides the weak eddy perpendicular scale 
\be
l_\bot\approx \lambda^{2/3} L^{1/3} M_A^{-2/3}.
\label{l_weak}
\ee

Unlike strong turbulence, the weak wave packets are cascading $\aleph$ times slower (see Eqs (\ref{tcas}), (\ref{aleph})), with
\be
\aleph \approx \left(\frac{V_A l_\bot}{u_l L}\right)^2,
\ee
where it is taken into account that the parallel scale of weak turbulence wavepackets is equal to the injection scale $L$. The rate of turbulent damping of
the Alfven wave is therefore
\be
\Gamma_{subA, w} \approx (\aleph \Delta t)^{-1}=\aleph^{-1}\frac{V_A}{L},
\label{gg}
\ee
which gives
\be
\Gamma_{subA, w} \approx \frac{V_A M_A^{8/3}}{\lambda^{2/3} L^{1/3}},
\label{gamma_weak}
\ee
which compared to the case of the earlier discussed damping  given by Eq. (\ref{gamma1}) shows even stronger dependence on $M_A$ 
as well as a different dependence of the wavelength $\lambda$. Being applicable to weak turbulence, this result does not transfer for $M_A=1$ to that in FG04 and therefore it is essential to define the range of its applicability in
terms of wavelength $\lambda$.   

 The maximal wavelength of the Alfven waves that can be cascaded by the
weak cascade can be obtained by substituting $l_{\bot}=L$, i.e. using the energy injection scale,
 in Eq. (\ref{lambda_w}). This gives:
\be
\lambda_{max, w}\approx L M_A.
\label{lambda_w_max}
\ee
Thus the particles emitting Alfven waves of the order of their gyroradius should have the range of gyroradii
\be
LM_A^4<r_L<LM_A
\label{range_weak}
\ee
in order to interact with weak turbulence. This is provided that $LM_A^4$ is larger than the damping scale of turbulent
motions. Otherwise the lower boundary in Eq. (\ref{range_weak}) is given by $l_{min}$. 

Waves with $\lambda>\lambda_{max, w}$ will interact with turbulence at the injection scale $L$. Such waves cascade 
by the largest wave packets whose cascading rate is $\aleph^{-1} \frac{V_A}{L}$, i.e. 
\be 
\Gamma_{outer}\approx \aleph^{-1} \frac{V_A}{L} \approx M_A^2\frac{V_A}{L},
\label{outer}
\ee
which is valid for $\lambda<L$. In the case of $\lambda\gg L$ the result in Eq. (\ref{outer}) is being reduced by another
random walk factor $(L/\lambda)^2$, i.e.
\be 
\Gamma_{outer, extreme}\approx \aleph^{-1} \frac{V_A}{L} \frac{L^2}{\lambda^2}\approx M_A^2\frac{V_A}{L}\frac{L^2}{\lambda^2}.
\label{outer_extreme}
\ee
which can be important for the damping of Alfvenic waves by turbulence injected at small scales. 

\subsection{Damping by SuperAlfvenic turbulence}

The case of superAlfvenic turbulence for scales less than the scale of the transfer to MHD regime, i.e. $l_{A}=LM_A^{-3}$, can be obtained from our
earlier results through the following considerations. At $l_{A}$ the turbulence becomes Alfvenic and this scale can be
considered as the turbulence injection scale. The injection velocity at this scale is $V_A$ and therefore the resulting
damping rate can be obtained by substituting $l_{A}$ as the injection scale $L$ and $V_L=V_A$ in Eq. (\ref{gamma1}). As a
result:
\be
\Gamma_{super}\approx \frac{V_A}{l_{A}^{1/2} \lambda^{1/2}}=\frac{V_A M_A^{3/2}}{L^{1/2}\lambda^{1/2}}.
\label{gamma_super}
\ee
 In a sense this is a case of transAlfvenic turbulence if $l_A$ is associated with the turbulence injection scale. This case corresponds 
to the FG04 where the turbulence injection scale was {\it defined} to be the scale $L_{MHD}$ at which the injection velocity becomes equal to $V_A$. Thus for superAlfvenic turbulence $L_{MHD}=l_A$.   

Treating $l_{A}$ as the effective injection scale one can easily get from Eq. (\ref{max1}) the maximal wavelength up to which
the above treatment of the non-linear damping is applicable:
\be
\lambda_{max, super}\approx l_{A}=LM_A^{-3}.
\label{max_super}
\ee
For the streaming instability we associate $\lambda$ with the gyroscale $r_L$ and therefore define the corresponding range 
gyroscales as
\be
\frac{l_{min}^{4/3}}{L^{1/3}} M_A<r_L<LM_A^{-3},
\label{range_super}
\ee
provided that $l_{min}<LM_A^{-3}$. In the opposite case of $l_{min}>l_{A}$ the turbulence is does not get Alfvenic over even at the smallest scales. 

For wavelengths larger than those given by Eq. (\ref{max_super}) and therefore for $r_L>LM_A^{-3}$ the damping is induced by Kolmogorov-type isotropic hydrodynamic turbulence which folds magnetic fields over the scale of eddies. The characteristic damping rate in this case is expected to coincide with the turnover time of the corresponding eddies, i.e.
\be
\Gamma_{hydro}\approx \frac{u_{\lambda}}{\lambda}\approx \frac{V_A M_A}{L^{1/3} \lambda^{2/3}},
\label{hydro_dam}
\ee
 where we used Eq. (\ref{u_hydro}).

\subsection{Other forms of presenting our results}

Emission of Alfven waves by energetic particles moving along magnetic field lines presents the most important case of the emission of Alfven waves in the local system of reference. The resonant emission along local magnetic field direction corresponds to the condition
\be
\lambda=r_L,
\label{l_r}
\ee
where $r_L=\gamma mc^2/eB$ is the Larmor radius of the resonant particle with a relativistic factor $\gamma$. We shall use Eq. (\ref{l_r}) in expressions below.

Expressing wave damping through the cascading rate is another way of presenting our results. 
Cascading of turbulent energy is a source of media heating. This can provide {\it upper limits} on the level of turbulence in astrophysical environments, which is valuable when the scales of the turbulent motions and injection rates are difficult to estimate.\footnote{The situation is changing with the development of new techniques that obtain the injection scale and injection velocity from observations (see Chepurnov et al. 2010, 2015, Lazarian \& Pogosyan 2012).} The cascading rate of the weak turbulence given by Eq. (\ref{eps_weak}) can be rewritten as
\be
\epsilon_w\approx \frac{V_A^3 M_A^4}{L},
\label{eps_w2}
\ee
which shows a decrease of the energy dissipation by a factor $M_A^4$ compared with the case of transAlfvenic turbulence. For $r_L<LM_A^4$ the damping rate for waves can be obtained by
expressing $M_A$ from Eq. (\ref{eps_w2}) and substituting it in Eq. (\ref{gamma1}):
\be
\Gamma_{subA, s}\approx \frac{\epsilon_w^{1/2}}{V_A^{1/2} r_L^{1/2}},
\label{gamma_subA_eps}
\ee
which differs from the expression in FG04 by the use of the cascading rate for weak turbulence $\epsilon_w$ instead of the cascading rate for strong turbulence. Thus the obtained damping rate for subAlfvenic turbulence is $M_A^2$ times less than in the case of trans-Alfvenic turbulence (see also Eq. (\ref{gamma1})).\footnote{It is interesting to note the special property of damping by strong turbulence. The damping depends only on the turbulent energy dissipation rate and not on the scale of the energy injection. Note, that the turbulent damping in other regimes is very different and does not show this remarkable universality.}  

For $L M_A^4<\lambda<LM_A$ we get the expression which is significantly different in its form from that in FG04. Indeed, expressing $M_A$ from Eq. (\ref{eps_w2}) and substituting it in Eq. (\ref{gamma_weak}) we can get
\be
\Gamma_{subA, w}\approx \frac{\epsilon_w^{1/3} L^{1/3}}{V_A r_L^{2/3}}\approx \frac{\epsilon^{1/3}_w M_A^{4/3}}{r_L^{2/3}}.
\label{gamma_weak2}
\ee
The expression given by Eq. (\ref{gamma_weak2}) has a slower dependence on the dissipation rate compared to Eq. (\ref{gamma_subA_eps}). The suppression of damping rate by the factor $M_A^{8/3}$ (see Eq. (\ref{gamma_weak})) is important and for $M_A\ll 1$ it explains the smooth transition to the regime of insignificant Alfven wave damping that is present for marginally perturbed magnetic fields. 

Dealing with the damping of Alfven waves emitted by particles with larger $r_L$, one can obtain the expression for the damping for $LM_A<r_L< L$ (see Eq. (\ref{lambda_w})) that corresponds to the damping by the outer scale of turbulent motions:
\be
\Gamma_{outer}\approx \frac{\epsilon_w^{1/2}}{L^{1/2} V_A^{1/2}}.
\label{outer}
\ee

For superAlfvenic strong MHD turbulence if one expresses $M_A$ from Eq. (\ref{super_alt1}) and substitutes it in Eq. (\ref{gamma_super}) it is easy to get
\be
\Gamma_{super}\approx \frac{\epsilon_{super}^{1/2}}{V_A^{1/2}r_L^{1/2}},
\ee
which has formally the same form as the expression for the damping for subAlfvenic strong turbulence given by Eq. (\ref{gamma_subA_eps}). The cardinal difference between the two expressions, assuming that the injection scale $L$ is the same, stems from the differences in the cascading rates in superAlfvenic and subAlfvenic turbulence. The subAlfvenic turbulence induces the significant {\it reduction} of the cascading rate compared to the transAlfvenic turbulence, the superAlfvenic strong MHD turbulence induces a significant {\it increase} of dissipation compared to the transAlfvenic case. 
Thus, for the same $L$, the damping of Alfven waves by superAlfvenic turbulence is more efficient than by the subAlfvenic one.
The damping rate for $r_L>\lambda_{max, super}$ where the latter is given by Eq. (\ref{max_super}) is produced by hydrodynamic turbulence and therefore is
\be
\Gamma_{hydro}\approx \frac{\epsilon_{hydro}^{1/3}}{r_L^{2/3}}.
\ee

In view of the astrophysical importance of damping in subAlfvenic turbulence, it is useful to rewrite the expressions given by Eq. (\ref{gamma1}) and (\ref{gamma_weak})  in terms of $\lambda_{max, s}$ given by Eq. (\ref{max_s}), namely
\be
\Gamma_{subA, s}\approx \frac{V_A}{L} \left(\frac{\lambda_{max, s}}{r_L}\right)^{1/2}, ~~~~r_L<\lambda_{max, s},
\label{g_s3}
\ee
and
\be
\Gamma_{subA, w}\approx \frac{V_A}{L} \left(\frac{\lambda_{max, s}}{r_L}\right)^{2/3}, ~~~~r_L>\lambda_{max, s}.
\label{g_w3}
\ee
Expressed in this form Eq. (\ref{g_s3}) explicitly shows that the damping by strong turbulence $\Gamma_{sub, s}$ is faster than the Alfven crossing rate of the injection scale eddies, while in the case of weak turbulence Eq. (\ref{g_w3}) shows that $\Gamma_{w}$ is slower that the aforementioned rate.

For $LM_A<r_L<L$, the damping rate can be written  as (see Eq. (\ref{outer}) and Eq. (\ref{max_s}))
\be
\Gamma_{outer} \approx \Gamma_{subA, s} \frac{r_L}{L},
\label{comp_outer}
\ee
which presents another form for the Alfven wave damping by turbulence at the outer scale.

\section{Damping of Alfven waves generated in the global system of reference}
 
Below we consider the damping of Alfven waves generated by an outside source which is not related to the magnetic field structure. It is important to realize that such waves should be viewed as being in the global system of reference and therefore our earlier treatment of the damping is not applicable. This is a separate case of damping relevant to many astrophysical settings, e.g. to the emission waves by stellar surface activity (see \S 6.5).  

\subsection{Case of Strong SubAlfvenic turbulence}

Consider first an Alfven wave moving at an angle $\theta\gg \delta B/B$ with respect to the {\it mean} magnetic field. In this situation one can disregard the dispersion of propagation angles that arises from turbulent magnetic wandering. For this purpose we use $\sin \theta$ instead of $\sin \theta_x$ in Eq. (\ref{alt_lambda}) and get for the perpendicular scale of eddies:
\be
x\approx \frac{\lambda}{\sin \theta}.
\label{xglobal}
\ee
The rest goes along the same line of reasoning that we employed in \S3.1. Indeed, the rate of the wave damping
is equal to the turnover rate of strong subAlfvenic eddies. Therefore using Eq. (\ref{xglobal}) it is easy to get
\be
\Gamma_{subA, s, \theta} \approx \frac{V_A M_A^{4/3} \sin^{2/3}\theta}{\lambda^{2/3} L^{1/3}},
\label{subA_s_global}
\ee
which provides the non-linear damping rate of an Alfven wave moving at the angle $\theta$ with respect to the mean field.

In terms of the cascading rate of weak turbulence $\epsilon_w$ (see Eq. (\ref{eps_weak})), the above damping rate for the wave
can be rewritten as:
\be
\Gamma_{subA, s, \theta} \approx \frac{\epsilon^{1/3}_w \sin^{2/3}\theta}{\lambda^{2/3}}.
\label{global_eps}
\ee

The turbulent damping given by Eq.(\ref{global_eps}) is applicable to 
\be
l_{min}\sin\theta < \lambda < LM_A^2 \sin\theta,
\label{range_new}
\ee
where $l_{min}$ is the minimal scale, i.e. the perpendicular damping scale, and $LM_A^2=l_{trans}$ is the 
maximal scale for the extent of the turbulent cascade. 

Naturally,  for this expression our approximation $\theta \gg \delta B/B$ fails if the wave is launched parallel to the mean magnetic field. The directions of 
the local magnetic field experience dispersion and this makes the actual $\theta_0$ not zero. In the global system of reference the
dispersion is determined by the magnetic field variations at the injection scale (see Cho et al. 2002). Therefore
\be
\theta_0\approx \frac{B_L}{B}\approx M_A.
\label{theta0}
\ee
Substituting this into Eq. (\ref{subA_s_global}) we get 
\be
\Gamma_{subA, s, 0}\approx \frac{\epsilon^{1/3}_w M_A^{2/3}}{\lambda^{2/3}},
\label{gamma_s_0}
\ee
which is different from our expression for the damping of Alfvenic waves moving along the local direction of the magnetic field (see Eqs. (\ref{gamma1}), (\ref{gamma_subA_eps})). The
difference stems from the difference in Alfven waves generated in respect to the local system of reference and in global system of
reference. 
The rate given by Eq. (\ref{gamma_s_0}) is applicable to the range
\be
l_{min} M_A < \lambda < LM_A^3,
\ee
which trivially follows from Eqs.(\ref{range_new}) and (\ref{theta0}).  

\subsection{The case of Weak SubAlfvenic turbulence}

For weak subAlfvenic turbulence in the case $\theta\gg \delta B/B$ we shall use Eq. (\ref{xglobal}) to relate the wavelength $\lambda$ to
the scale of perpendicular motions that the wave interacts with while cascading as well as Eq. (\ref{gg}) to get the damping rate corresponding to such
motions. As a result,
\be
\Gamma_{weak, global, \theta} \approx \frac{V_A \sin\theta M_A^2}{\lambda}\approx \frac{\epsilon^{1/2} L^{1/2} \sin\theta}{V_A^{3/2} \lambda},
\label{global_weak}
\ee
where in the damping is expressed through the weak cascading rate $\epsilon_w$.

The applicability of this type of damping is applicable to 
\be
LM_A^2 \sin\theta< \lambda < LM_A \sin\theta,
\label{range_new2}
\ee
where the last inequality is obtained by substituting the maximal perpendicular scale of eddies $LM_A$ for $x$ in 
Eq.(\ref{xglobal}).

For the propagation along the mean magnetic field one should take into account Eq. (\ref{theta0}) which results in
\be
\Gamma_{weak, global, 0}\approx \frac{V_A M_A^3}{\lambda}\approx \frac{V_A \epsilon^{3/4} L^{3/4}}{\lambda V_A^{5/4}}.
\label{global_weak_0}
\ee
The range of the applicability of this damping rate is 
\be
LM_A^3 < \lambda <LM_A^2,
\ee
where Eq.(\ref{theta0}) and (\ref{range_new2}) were used.

 \subsection{Other cases}
 
 For strong superAlfvenic turbulence, i.e. for damping by turbulent motions at scales less than $l_A$ one
 can still use our approach above and consider damping of Alfven waves with $\lambda< l_A$ (see Eq. (\ref{max_super})). The damping by 
 eddies less than $l_A$ happens by one eddy turnover time. If the wave is at an angle $\theta$ to the magnetic field within a magnetic eddy $<l_A$ then the
 damping happens over one turnover time for the motions of the size $x$ defined by Eq. (\ref{xglobal}). The procedures
 analogous to those we employed above provide
 \be
 \Gamma_{super, global, \theta}\approx \frac{V_A M_A \sin^{2/3}\theta}{\lambda^{2/3} L^{1/3}},
 \label{global_super}
 \ee
 where in superAlfvenic turbulence angle $\theta$ changes from one strong turbulence eddy of size $l_A$ to another. Therefore an averaging over such changing directions should be performed which for the random distribution of directions provides the damping rate of $\langle \sin^{2/3} \theta \rangle =3/5$.
 
 For Alfven waves from a macroscopic source $\gg l_A$ the turbulent volume can be considered as consisting of MHD cells with the regular MHD turbulence 
 but with the injection of transAlfvenic turbulence at the scale $l_A$. The wave damping will differ depending on the angle $\theta$ between the magnetic field in a
 given cell and the wave propagation direction. The rate of damping can be obtained by substituting in Eq. (\ref{subA_s_global}) the actual angle $\theta$ as well as $M_A=1$ and $L=l_A$. The minimal wavelength in this case depends on the $l_{min}\sim \theta$. 
    
 At scales larger than $l_A$ the turbulence is essentially hydrodynamic. Therefore, for turbulent damping by superAlfvenic eddies of size larger than $l_A$ as well as for damping by outer-scale eddies there is no difference between local and global frames. Therefore our earlier results are applicable. 
 
 \subsection{Finite-sized macroscopic emitter}
 
 Our considerations obtained for an infinitely extended microscopic emitter can be generalized for the finite size emitter. If the size of the emitter $y$ and the wave is emitted at the angle $\theta\gg \delta B_{*}/B$, then our considerations in \S 5.1-5.2 stay the same. Note, however, that $B_{*}$ in this case is $min[B_L, B_{damp}]$, where $B_{damp}$ is the magnetic field deviation at the scale of wave damping, i.e. $l_{damp}\approx \Gamma_{global, \theta}^{-1} V_A$, where $\Gamma_{global, \theta}$ are, for instance, defined for weak and strong subAlfvenic turbulence in \S 5.1 and \S 5.2. 
 
 If the wave is emitted parallel to the local magnetic field at the scale $y$, then we are have to deal with the intermediate case having features of Alfven wave damping in local and global systems of reference. Indeed, the variations of the magnetic field directions should be calculated at the scale of $y$ and compared with the variations of magnetic field at the "resonant" scale.
 For strong subAlfvenic turbulence this scale is given by Eq. (\ref{x1}) and for weak subAlfvenic turbulence by Eq. (\ref{l_weak}).
 The latter two scales depend on $\lambda$. Therefore we expect to see the scaling corresponding to Alfven wave damping if the
 $y$ is smaller than the values given by the aforementioned equations. Note that the damping of the waves emitting with respect to
 the mean magnetic field will be happening inhomogeneously with patches where the local magnetic field happens to be parallel
 to the wavefront having the ability to support the Alfven wave propagation for a longer period of time.

 \section{Comparison with non-linear Landau damping}
 
 It is important to compare the turbulent damping that we study in this paper with the non-linear Landau damping process that can also damp Alfven waves (see Kulsrud 2005).  The latter damping is inversely proportional to the square root of CR scaleheight $L_z$, so we may expect that this process is subdominant for weak gradients of the cosmic ray distribution. Indeed,  the ratio of the rate of turbulent subAlfvenic damping and the rate non-linear Landau damping $G_{NL}$ can be evaluated to give:
\be
\frac{\Gamma_{subA, s}}{\Gamma_{NL}} \approx  \frac{B_{\rm \mu G}^{3/2} n_{i,-3}^{1/4} L_{\rm z,100}^{1/2} M_A^2}{L_{\rm 100}^{1/2}T_{4keV}^{1/4} n_{\rm CR,-10}^{1/2}} \gamma_{100}^{n/2-2},
\label{ratio_damp}
\ee
where $T_{\rm 4 \, keV}=(T/4 \, {\rm keV})$, $B_{\rm \mu G}=(B/1 \, \mu G)$, $L_{\rm z,100}=(L_{z}/100 \, {\rm kpc})$, $n^{\rm i}_{-3}=(n_{\rm i}/10^{-3} \, {\rm cm^{-3}})$, $n^{\rm CR}_{-10}=n^{CR}(\gamma > 1)/10^{-10} \, {\rm cm^{-3}})$, $\gamma_{100}=\gamma/100$, and is scaled to $n=4.6$. Note that $n^{CR} (> \gamma) = 10^{-10} \gamma^{-1.6} \, {\rm cm^{-3}}$ of the order a CR energy density in equipartition with a $\sim \mu$G B-field.
 Therefore, if  the CR profile falls ($n_{\rm CR} \rightarrow 0$) and flattens ($L_{\rm z} \rightarrow \infty$) the non-linear Landau damping becomes subdominant.
 
 One may wonder whether  for very weak levels of turbulence $M_A\rightarrow 0$ the non-linear
Landau damping may become important.  The latter, however, is a self-regulated process as the suppression of the streaming instability is bound to allow the CRs to spread fast, decreasing the CR gradient. Potentially resonance scattering could mitigate this spreading. However, this depends on the presence of fast modes that, in the absence of streaming instabilities, were identified in Yan \& Lazarian (2002) as a major factor of cosmic ray scattering in the interstellar plasmas..\footnote{The importance of fast waves is easy to understand. One should recall  that  that due to the extreme anisotropy of the tensor that describe the Alfven turbulence at small scales (Cho et al. 2002), the scattering by Alfvenic modes of the MHD cascade initiated at the large injection scale $L$ is very small (Chandran 2000, Yan \& Lazarian 2002).}  In many instancies, e.g. galactic halos, these modes are subject to significant collisionless damping and therefore their efficiency of controlling of the CR spreading is limited.
On the contrary, there is no such a self-regulation for
turbulent damping of CR streaming which makes the process dominant in most astrophysical settings. In many instances when the turbulent damping fails, the non-linear Landau damping is unlikely to damp the CR streaming either. For instance, we argue in the next section that the turbulent damping of the streaming instability is not important for the Milky Way halo due to the low level of turbulence there. There we
do not expect non-linear damping to be important there due to the self-regulation which entails the increase of $L_{\rm z}$.

\section{Astrophysical Implications}

In what follows we discuss in detail the problem of damping of streaming instability in our Galaxy and sketch some other astrophysical implications of the improved understanding of Alfven wave damping that we have obtained in this paper. The detailed treatment of these implications will be provided elsewhere. 

\subsection{Streaming of CRs in galaxies}

One of the simplest models of the galactic CR propagation is the so-called "leaky box model"  (see Longair 2011). Within this model CRs propagate freely within the galactic disk, while they experience streaming instability as they enter a fully ionized halo surrounding the galaxy. Free zooming through the galactic disk is possible as, in the leaky box model, the galactic disk is assumed to be partially ionized and therefore the streaming instability is being suppressed by ion-neutral damping (see Kulsrund \& Pearce 1968). This model is surely naive, as the galactic disk is definitely not fully filled with partially ionized gas. In fact, a significant fraction of the the galactic disk is filled with hot ionized gas (McKee \& Ostriker 1977, see Draine 2011). Moreover, the leaky model does not account for turbulent damping of streaming instability. 

The first treatment of CR propagation that took the streaming instability damping into account was done by FG04. This study came to a paradoxical conclusion, namely, that turbulence suppresses streaming instability for most of the CR energies and therefore it is really difficult to understand the observed high isotropy of CRs. Below we subject the the problem to scrutiny and come to the conclusions that are different in FG04. In particular, we will show that (a) damping is produced by weak rather than strong Alfvenic turbulence and therefore is reduced, (b) the turbulent dissipation rate assumed in FG04 to be equal to the plasma cooling rate  is overestimate of the actual dissipation rate, as this way of estimating disregards other important heating mechanisms. 

Our study above shows that turbulent damping of the streaming instability changes significantly with whether the damping is performed by strong or weak Alfvenic turbulence. Note, that it is natural to assume from the very beginning that turbulence in the halo is subAlfvenic rather transAflvenic or superAlfvenic. This fact is easy to understand. Indeed, the ISM is turbulent (see Armstrong et al. 1995, Elmegreen \& Scalo 2004, McKee \& Ostriker 2007, Chepurnov \& Lazarian 2010) and the sources of turbulence driving, whether they are related to supernovae (see MacLow 2004, Draine 2011) or magnetorotational instability (see MacLow \& Klessen 2004), are within the galactic disk. The magnetic field in the halo is expected to become more and more quiescent with the greater distance from the disk as turbulence decays diffusing from the disk. Our quantitative estimates based on the observational data that we provide below support this intuitive notion.  

For subAlfvenic turbulence it is possible to express the streaming rate using the textbook approach to the streaming instability (see Kulsrud 2005), but equating the turbulent damping rate to the streaming damping rate in Eq. (\ref{gamm_cr})  (see FG04):
\be
v_{stream}\approx V_A \left(1+\frac{\Gamma n_i r_L}{\gamma c n_{cr}}\right),
\label{gen_st}
\ee
where we used the relation $r_L=\gamma c \Omega^{-1}$. Note, that the rates of damping $\Gamma$ are different for weak and strong turbulence. In particular, the damping is
by strong turbulence if $\Gamma=\Gamma_{sub, s}$, i.e. for $r_L<LM_A^4$, and  by weak turbulence if $\Gamma=\Gamma_{w}$, i.e. for $r_L>LM_A^4$. For the Milky Way galaxy the quantities that enter Eq. (\ref{gen_st}) can be estimated for the hot coronal gas of the halo, i.e. plasma with density $n_i\approx 10^{-3}$ cm$^{-3}$ temperature $T\approx 10^{6}$ K  and magnetic field $B\approx 3$ $\mu$G. An upper limit of the Alfven Mach number $M_A$ may be obtained for the galactic halo assuming that the turbulent velocity dispersion in the halo is the same as in the disk, i.e. $10^6$ cm/s. Indeed, as the sources of the turbulence are localized in the galactic disk and the turbulence decays quickly (see Stone et al. 1998, Cho \& Lazarian 2002), this value is a substantial overestimate of the turbulent velocities. With the Alfven velocity for the parameters above being $V_A\approx 2\times 10^7$ cm/s, one gets $M_A<1/20$. Therefore the critical value of $r_L$ is $LM_A^4\approx 5\times 10^{14}$ cm, where we assumed the injection scale $L$ equal to 100 pc. The relativistic proton gyroradius is $r_L\approx 10^{12} \gamma$ cm, which means that the streaming that is controlled by strong turbulence is applicable up to 
$\gamma< 500$. This conclusion contradicts to the use of damping by strong turbulence assumed in FG04. In fact, we believe that the expression of streaming affected by strong turbulence, namely, 
\be
v_{stream, s}\approx V_A \left( 1+ \left[\frac{\epsilon_w}{700 erg~s^{-1}~g^{-1}}\right]^{1/2} \gamma^{1.1}\right),
\label{streaming_strong}
 \ee
 is applicable to the damping of turbulence in the galactic disk rather to any parts of the Milky Way halo. 
Note, that Eq. (\ref{streaming_strong}) in its form coincides with the expression for the streaming velocity in FG04. This coincidence is the result of the remarkable universality of the damping by strong Alfvenic turbulence that we discussed earlier in \S 4.5. However, the significant difference of Eq. (\ref{streaming_strong}) expression is the use of the weak turbulence cascading rate $\epsilon_w$. This rate differs from the one for transAlfvenic turbulence by a factor $M_A^4$. 

 Most of the streaming cosmic rays in the Milky Way halo are expected to have $r_L>LM_A^4$. For them, using the expression for damping by weak turbulence, i.e. Eq. (\ref{gamma_weak2}), it easy to obtain
\be
v_{stream, w}\approx V_A \left(1+\frac{\epsilon^{1/3}_w n_i r_L^{1/3} M_A^{4/3}}{\gamma c n_{cr}}\right).
\label{st_w}
\ee 
It is evident that Eq. (\ref{st_w}) is very different from Eq. (\ref{streaming_strong}). The most significant difference stems from the fact that the streaming velocity in Eq. (\ref{st_w}) does depend not only on the dissipation rate 
$\epsilon_w$, but also on the Alfven Mach number $M_A$. It is also important  that
the damping rate in Eq. (\ref{st_w}) depend on the $\epsilon_w^{1/3}$ rather than $\epsilon_w^{1/2}$ as in Eq. (\ref{streaming_strong}). Indeed, both factors above help avoiding "streaming catastrophe" outlined in FG04.  
Additional, but less important factor that helps us is that the second term of Eq. (\ref{st_w}) scales as $\gamma^{0.94}$ compared to $\gamma^{1.1}$ in Eq. (\ref{streaming_strong}). 
 
To find the streaming velocity using Eq. (\ref{st_w}) one should know both the dissipation of weak turbulence
$\epsilon_w$ and the Alfven Mach number $M_A$. Two different estimates of the cascading rate were presented in FG04. One was based on the cooling rate for the hot gas, the other was based on the supernovae energy injection rate. The latter is readily available. Indeed, it is accepted that the supernovae are releasing $10^{51}$ ergs of mechanical energy in the gas once every one million years in the disk area of 100 pc$^2$ (see Draine 2011). Assuming that the resulting turbulence is transAlfvenic and therefore decays in one Alfven crossing time the FG04 obtained the estimates for the turbulent dissipation rate of $\approx 25$ erg s$^{-1}$ g$^{-1}$. We believe that this is an estimate that has relevance to the galactic disk, rather than to the galactic halo. Such a significant rate of turbulent dissipation according to Eq. (\ref{streaming_strong}) should ensure that within the media of the galactic disk  Alfvenic turbulence suppresses streaming instability, which corresponds to the disk part of the "leaky box" model. 

The situation is very different for the hot plasmas in the Milky Way halo. There the second estimate in FG04 based on the gas cooling might be relevant. Indeed, the turbulent cascading rate determines hot gas heating and this cannot be larger that the radiative cooling rate, which is about $0. 06$ erg s$^{-1}$ g$^{-1}$ (see Binney \& Tremaine 1987). In fact, this provides the {\it upper limit} for the turbulent cascading and the actual rate, as we discuss further, may be substantially lower.\footnote{The rate of turbulent heating by supernovae above and the {\it upper limit} of turbulent heating at hand are so different both due to the decrease of turbulent velocities in the galactic halo compared to the disk and to the decrease of Alfven Mach number $M_A$. The latter makes turbulence less dissipative in proportion to   $M_A$.} Indeed, turbulent heating is not the only way of heating the halo plasmas. For instance, we may consider heating that comes from CR streaming (see Wiener et al. 2013b).  The irreversible energy transfer from streaming CRs to gas provides the volumetric heating rate (see Kulsrud 2005):
\be
\Gamma_{heat}\approx -V_A \bigtriangledown P_{cr}.
\label{heat}
\ee
To get the heating per unit of mass one has to divide the heating rate given by Eq. (\ref{heat}) by the plasma density. Taking as a rough estimate the energy density of CRs to be 1 eV per cm$^{-3}$ and the characteristic scale of the CR change to be $L_{cr}\approx 5$ kpc, one gets heating $\sim 0.06$ erg s$^{-1}$ g$^{-1}$, which coincides with the cooling rate in Binney \& Tremaine (1987). This may indicate that the galactic halo is heated by cosmic ray streaming that {\it does take place} in the halo environment. As a result, the cascading rate adopted in FG04 significantly {\it overestimates} the actual turbulence cascading in the halo gas of the Milky Way\footnote{We note parenthetically that the adopted cascading of $\sim 0.06$ erg s$^{-1}$ g$^{-1}$ corresponds to $M_A\approx 0.2$ if the injection scale $L=100$ pc is adopted. This suggests that even with this cascading rate that we argue to be an overestimate, the turbulence is subAlfvenic.}. We feel that the most important is to establish whether streaming instability really fails in the realistically turbulent Milky Way halo. Therefore, for the rest of our discussion, we concentrate on showing that the conclusions about the "streaming catastrophe" reached in FG04 are not obtained on the self-consistent basis. 

It is well known that the anisotropy is less than $0.1\%$ for the CRs with $\gamma<10^6$ (see Longair 2011).  As the Alfven velocity in the hot plasmas is $\sim 0.1\%$ of $c$ FG04 assumed that the the second term in brackets of Eq. (\ref{streaming_strong}) is not larger than unity.\footnote{The difference in terms of strong turbulence cascading assumed in FG04 and the weak that is employed in Eq. (\ref{streaming_strong}) is not important for the argument as we discussed earlier.} This provided $V_{streaming}\approx V_A (1+0.01\gamma^{1.1})$ for the cascading of $\sim 0.06$ erg s$^{-1}$ g$^{-1}$. On the basis of this estimate, FG04 concluded that to avoid the contradiction with the observational data for $\gamma\sim 10^6$ one should assume that the rate for the turbulent dissipation is less than $4 \times 10^{-11}$  erg s$^{-1}$ g$^{-1}$, which is very different from the assumed  $\sim 0.06$ erg s$^{-1}$ g$^{-1}$ rate. On the basis of this FG06 came to the conclusion that streaming instability is not feasible as the solution for solving the problem of explaining the observed isotropy of cosmic rays. 

As we pointed above, for realistic magnetization of the galactic halo $M_A\ll 1$ and Eq. (\ref{st_w}) rather than
Eq. (\ref{streaming_strong}) should be used to determine the streaming velocities.\footnote{In fact, for $M_A<0.003$ the CR with $\gamma=10^6$ interact with the turbulence in the outer scale, which further reduces turbulent damping (see Eq. (\ref{outer}).}  It is safe to say that in the situation the turbulent damping in galactic halo not being constrained observationally it is premature to be alarmed about the failure of streaming instability to explain the cosmic ray isotropy are premature. In fact, we expect the turbulent velocity to decrease fast with the distance from the galactic plane. In addition, due to the drop of the plasma density, we also expect the exponential increase of $V_A$ with the distance from the galactic plane. Therefore in Eq. (\ref{st_w}) both $\epsilon_w$ and $M_A$ are likely to decrease exponentially, i.e. $\sim \exp (-H/h)$, where $H$ is the halo size $\sim 5$ kpc, and $h\sim L$ is the scale height of
the galactic disk, which is one or two hundred parsecs. Therefore it is likely that at {\it some} distance from the disk $\gg L$ the second term in brackets in Eq. (\ref{st_w}) becomes small. This is what the only thing that is required for the streaming instability to isotropize CRs. 

Expressing the streaming velocity through the turbulence dissipation rate is advantageous only when this dissipation rate is readily available from observations. In the situation of galactic halo when the turbulent heating may not be the dominant process it seems advantageous to use the other expressions for the turbulent damping rate, e.g. for the damping by weak turbulence to substitute in Eq. (\ref{gen_st}) the expression for damping given by Eq. (\ref{gamma_weak}). This way we get:
\be
v_{stream, w}\approx V_A \left(1+\frac{V_A n_i r_L^{1/3} M_A^{8/3}}{L^{1/3}\gamma c n_{cr}}\right),
\label{stream_w_alt}
\ee
where the turbulence injection scale $L$ can be obtained from observations with statistical techniques using spectral lines (see Chepurnov et al. 2010, 2015) or synchrotron emission (see Lazarian \& Pogosyan 2012, 2016), while the Alfven Mach number $M_A$ can be obtained using anisotropy studies with spectral lines (see Esquivel \& Lazarian 2005, Burkhart et al. 2014, Esquivel, Lazarian \& Pogosyan 2015, Kandel, Lazarian \& Pogosyan 2016ab) or synchrotron studies (see Lazarian \& Pogosyan 2012, 2016, Herron et al. 2016). In particular the variations of $L$ and $M_A$ with the distance from the observer can be obtained using multifrequency polarization studies as explained in Lazarian \& Pogosyan (2016). We believe that this is a promising future direction of research.

In fact, in view of our study, the "leaky box" model can be reformulated. Instead of suppression of streaming instability in the disk by ion-neutral collisions, the instability is likely to be efficiently suppressed by turbulence there. At the same time, the streaming instability can be present in the Milky Way halo, returning and isotropizing CRs. 

Naturally, apart from the streaming instability, there are also other ways to isotropize CRs. We also note that additional sources of CR isotropization come from the scattering of CRs as well as from magnetic field wandering\footnote{Magnetic field wandering for Alfvenic turbulence was first described in LV99 and later employed in solving different problems from thermal conduction of magnetized plasmas (Narayan \& Medvedev 2002, Lazarian 2006) to shock acceleration (Lazarian \& Yan 2014).}. Due to the Richardson dispersion (see Lazarian \& Yan 2014) CRs following magnetic field lines spread superbalistically in the direction perpendicular to the mean magnetic field, modifying and decreasing the anisotropies (see Lopez-Barquero et al. 2015). In addition, fast modes that were identified as the major source of scattering in the galactic environments in Yan \& Lazarian (2002, 2004) can provide significant isotropization of CR. These possibilities were not considered in FG04. On the contrary, the idea that is mentioned there, i.e. of confinement using magnetic mirror arising from dense molecular clouds (Chandran 2000) looks problematic. Indeed, in view of the low filling factor of dense clouds it looks unrealistic to think that CRs have to encounter many magnetic bottles created this way prior to  their leaving the galaxy. In addition, with the new data that shows that the strength of magnetic fields stay in a significant fraction of molecular clouds on the level close to the value of the field in diffuse interstellar medium  (Crutcher et al. 2010), the confinement efficiency of magnetic bottles created by molecular clouds  is very questionable\footnote{The effect of poor correlation of density and magnetic field was explained in Lazarian, Esquivel \& Crutcher (2012) as the consequence of process of turbulent reconnection or "reconnection diffusion" (LV99, Lazarian 2005, Santos-Lima et al. 2010).}. Moreover, the formation of magnetic bottles does not ensure particle anisotropy, as the magnetic bottles formed by molecular clouds are stationary and therefore they do not
change the adiabatic invariant of the particles confined by the bottles.

\subsection{Acceleration of particles in shocks and reconnection layers, impact on reconnection}

We also want to stress that the issues related to CR streaming are not limited to the observed CR isotropy. We believe that the CR streaming instability can be present also in the galactic disk but at places of significantly higher than average CR flux, e.g.  near places of CR acceleration, e.g. shocks (see Bell 1978, Schlickeiser 2002) or reconnection sites (de Gouveia dal Pino \& Lazarain 2005, Lazarian 2005, Drake et al. 2006, Lazarian \& Opher 2008).

The acceleration of CRs in shocks is an accepted process for explaining the population of galactic CRs (Krymski et al. 1978, Bell 1978, Armstrong \& Decker 1979). To be efficient, the process of returning of CRs back to the shock must also be efficient. Potentially, the streaming instability should be important for returning particles back (see Longair 2011). Turbulence, however, is likely to complicate the process. In fact, apart from the pre-existing turbulence, there is turbulence that is generated both in the precursor (Beresnyak et al. 2009, del Vale et al. 2016) and the postshock media (Giacalone \& Jokipii 2007). This superAlfvenic small-scale turbulence is expected to efficiently damp the streaming. At the same time, the same turbulence also generates a turbulent magnetic field, which can act as a magnetic mirror that returns the CRs back to the shock. Therefore, it is likely
that the CR acceleration in shocks proceeds without the important contribution from the streaming instability.

Streaming instability can also return particles accelerated by magnetic reconnection to the reconnection site enhancing the First order Fermi acceleration that arises from reconnection (de Gouveia dal Pino \& Lazarian 2005,
Drake et al. 2006). The corresponding reconnection can proceed both when large scale magnetic field reconnects releasing its free energy and within multiple reconnection regions in the steady state turbulence. The latter process was recently considered in Brunetti \& Lazarian (2016). The role of the streaming instability depends on the level of turbulence in the system. Generically, we expect the level of turbulence to increase in the reconnection regions as magnetic reconnection progresses (see Lazarian et al. 2016) and therefore the role of streaming instability to decrease. However, the study of the parameter space for which the streaming is important both for magnetic reconnection and shock CR acceleration is beyond the scope of the present study. 

We also note that magnetic reconnection can be a source of Alfvenic waves (see Kigure et al. 2010).   As the process of reconnection happens generically in turbulent fluids, it is natural that the 
generated Alfven waves should experience turbulent damping. Eventually, as we discussed, this should contribute to generating more turbulence in the reconnection region. Turbulence was shown in LV99 to change the nature of magnetic reconnection making it independent of plasma resistivity (see more in Kowal et al. 2009, 2012, Eyink, Lazarian, Vishniac 2011, Eyink et al. 2013, Eyink 2015, Lalescu et al. 2015). Turbulence is being generated by  reconnection thus inducing fast reconnection in the case when the initial state of magnetized plasmas is not turbulent (Beresnyak 2013,  Oishi et al. 2015, Lazarian et al. 2015). In highly magnetized plasmas with magnetic energy significantly exceeding thermal energy, the transition to turbulent reconnection has an explosive character with higher level of turbulence increasing the rate of reconnection and the higher reconnection increasing the level of turbulence (LV99, Lazarian \& Vishniac 2009). Our study shows that the transition to turbulence is inevitable even if initially a significant part of energy leaves the reconnection zone in the form of Alfven waves. 

\subsection{Implications for galaxy clusters}

In WOG (see also Ensslin et al. 2011, Pinzke et al. 2015) streaming instability suppression was invoked to explain the bimodality of the cluster radio emission, namely, the fact that the majority of clusters are radio-quiet (Brunetti et al. 2007, 2009, Brown et al. 2011, Brunetti \& Jones 2014 and ref. therein), and it is only the clusters associated with merger activity that demonstrate radio halos. The authors above suggested a way to account for this property by assuming that  the CRs escape at  superAlfvenic speeds  and this fast escape turns off  the radio galaxies (see Ensslin et al. 2011). 

It was shown in WOG that  non-linear Landau damping (e.g. Felice \& Kulsrud 2001) is too weak to inhibit wave growth, while turbulent damping (YL02, FG04) can suppress the instability. This conclusion agrees with our analysis in \S 6. Moreover, our present study allows us to express the results in WOG in terms
of the actual parameters of the turbulence in galaxy clusters, e.g. their magnetization and the turbulence injection scale.  This turbulence is accepted to be superAlfvenic (see Brunetti \& Lazarian 2007, Miniatti \& Beresnyak 2015).  The Alfven Mach number of the intracluster medium is expected to vary depending on the level of turbulence. In Brunetti \& Lazarian (2016) the range of $M_A$ was estimated to be from 3 to 9. The value of $l_A$ thus may range from approximately 10 pc to 0.3 pc if we assume the injection scale of $10^2$ pc. Our study dictates that these values of $l_A$ should be used in WOG for $L_{MHD}$ that they employ in their study while dealing with the streaming instability by strong MHD turbulence. 
Adopting $l_A=r_L=1$ pc one gets that CR with $\gamma<10^6$ interact with strong turbulence, as it is assumed in WOG. At the same time the streaming induced by CR with higher 
$\gamma$ is affected by the damping induced by superAlfvenic turbulence in hydrodynamic regime. 
 
 Our quantitative insight  strengthen the conclusion in WOG that the CRs can stream rapidly in the presence of superAlfvenic turbulence. However, the consequences of this effect for the dynamics of CRs on large scales are not easy to evaluate. Indeed, the escape of CRs is limited not only by streaming but also by turbulence scattering as well as the diffusion of magnetic field lines. The latter in superAlfvenic turbulence are entangled on the scale $l_A$ (e.g. Brunetti \& Lazarian 2007), which produces the random walk with the scale of $l_A$. This entails the increase of the escape time by a factor $(D/l_A)^2$, where $D$ is the length of order of Mpc that the particles should cover, while our estimate of $l_A$ is of the order of 1pc. These are the complications that should be considered in the future quantitative models. The process of turning off and on can also be explained by merger-induced scenarios of turbulent reacceleration as discussed in detail by Brunetti \& Jones (2014 and ref. therein). A synthesis of the approaches above will be presented in a future publication.

\subsection{Streaming of CRs and ionization of molecular clouds}

Streaming of CRs into molecular clouds is an interesting process that requires further studies. For instance, in a recent paper by  Schlickeiser et al. (2016) 
streaming instability arising as the CRs penetrate molecular clouds was described. This study, however, does not account for a possible suppression of streaming instability by ambient turbulence. For superAlfvenic turbulence, the processes of penetration of CR inside the clouds are going to be modified.
As a result one can imagine a situation in which the coefficient for the "along the magnetic field" diffusion is larger in the outer turbulent parts of the molecular cloud and smaller at the inner part of the molecular cloud where the streaming instability operates and creates waves scattering CRs. In this situation the density of CRs may potentially be higher in the interior of molecular clouds than in the ambient interstellar medium. This can also be relevant to explaining observations (see McCall et al. 2003, Le Petit et al. 2004) which suggest significant variations of the CR density in molecular gas. In realistic inhomogeneous interstellar gas one can expect regions where streaming instability is suppressed and regions where it operates, creating significant variations of  CR diffusion and CR density. 

\subsection{Heating of plasmas and launching of winds}

While the damping of Alfven waves by turbulence has become a well accepted process in the field of CR research, in other fields the studies of Alfven damping frequently ignore the turbulent nature of the magnetized plasmas and focus instead of wave steepening and pure plasma effects. Therefore we would like to point out that our results are applicable to heating of stellar corona by Alfven waves and launching of stellar winds by damping of Alfven waves (see Suzuki \& Inutsuka 2005, Verdini et al. 2005, Evans et al. 2009, Vidotto \& Jatenco-Pereira 2010, Verdini et al. 2010, Suzuki 2015). 
The cascading that we consider results in efficient dissipation of Alfven waves and this dissipation is very robust, i.e.
it does not depend on the microphysics of plasma processes. Our results show that in highly magnetized regions of solar atmosphere with low 
Alfven Mach number $M_A$ Alfven waves can propagate larger distances than in regions with lower $M_A$. This should be accounted in the quantiative modeling of wind launching and plasma heating. We note that the turbulent damping scenario does not require
efficient coupling between Alfven and fast mode turbulence that is assumed in some of the studies (see Cramer et al. 2014), it does not require having non-linear Alfven waves of large amplitude either (cf. Airapetian et al. 2010). For instance, we believe that the turbulent damping
can be relevant to explaining the observed "unexpected" damping of Alfven waves in the regions above the Sun's polar coronal holes (Hahn et al. 2012). These and other issues should be clarified by the further research which accounts for the turbulent damping of Alfven waves.

Heating by waves emitted by various sources can be an important source of heating of turbulent plasmas in galaxy clusters. Our study provides a way to quantify the distribution of heating as a function of the distance from the source.

Heating by Alfven waves emitted by processes on the stellar surface and the processes of launching of stellar winds are intrinsically connected. Alfven waves emitted by stars carry momentum. This momentum is deposited with the plasmas as Alfven waves dissipate and this can be the process that launches the wind itself or contributes to the process of the wind launching, e.g. together with the radiation force  (see Suzuki 2011, 2015). The efficient turbulent damping of Alfven waves that we have demonstrated in this paper makes this process efficient. 

Interestingly enough, a sufficiently strong flux of Alfven waves can induce an instability, resulting in the formation of the area of enhanced turbulence damping.  Consider a train of Alfven waves subject to turbulent damping in magnetized plasmas where a particular region has an enhanced level of turbulence. This area will induce stronger damping of Alfven waves. Those, as they cascade, will decrease their perpendicular scale until their parallel and perpendicular scales eventually satisfy Eq. (\ref{crit}) corresponding to the turbulence critical balance. So the cascading Alfven waves will create more turbulence which through inverse cascading can produce motions that can cascade the  Alfven waves more efficiently.\footnote{This turbulence can
scatter fluctuations of smaller wavelength than the original wavelength of the train. It can also transfer some part of the energy to large scales through the inverse cascading process and thus increase the turbulent damping of the original train of Alfven waves. The difference between the scales at which the damping of waves 
occurs depends on the angle $\theta$ between the direction of Alfven waves and the mean magnetic field as well as Alfven Mach number of the original turbulence.
For sufficiently large $\theta$ or/and sufficiently large $M_A$ the scales of turbulent damping and the transfer of of the energy into the energy of turbulent motions can be close making the instability efficient.}  The turbulence initially gets imbalanced, but in realistic turbulent media with
density inhomogeneities as well as in the presence of parametric instabilities  (del Zanna et al. 2001) the scattered waves become a part of the balanced MHD turbulence.  

The dissipation of Alfven waves that we discussed above happens in the global system of reference.
However this does not exhaust all the possibilities for launching the winds. For instance, CR can launch winds getting coupled with the magnetized plasmas through streaming instability (see Recchia et al. 2016). In the latter case, turbulent damping of Alfven waves happens in the local system of reference. Our work shows that quantitative models of CR-driven winds (see Ruskowski et al. 2016) should account for the spatial change of turbulent damping arising from the change of $M_A$. We expect to see near the galactic disk the superAlfvenic CR streaming that was invoked by Ruskowski et al. (2016) in their modeling. This, however, should change to Alfvenic streaming in the galactic halo as the turbulent damping is expected to get less efficient there. The consequences of this change are difficult to evaluate without detailed calculations.

\section{Comparison with earlier works}

Silimon \& Sudan (1989) for their studies of Alfven waves damping
used models of MHD turbulence that were not supported by further research.
Yan \& Lazarian (2002) suggested that the streaming instability can be suppressed by 
turbulence, but did not provide a quantitative study of the processes. In this situation
the closest study to the present one is the quantitative pioneering study of the streaming instability 
damping by strong Alfvenic turbulence in FG04. In view of the theory provided in this paper this is a particular case of damping.
 As we identified in our paper, this is the case of the Alfven waves that are emitted 
in the local system of reference.  In the present paper we also identified the other regime
of damping, i.e. when the damping of Alfven waves emitted
by a macroscopic source. The latter damping happens with respect to the mean magnetic field, i.e. in the global system of reference.
The scalings of the damping are different in two cases (see Table 1).

As for the streaming instability damping, our treatment is different from FG04 and we explain the differences below.
The model of turbulence adopted in FG04 is based on the assumption that the turbulent energy is injected at the
scale $L_{MHD}$ with velocity $V_A$. This is a case of transAlfenic turbulence with the caveat that FG04
does not associate $L_{MHD}$ directly with the injection scale, but {\it defines} this scale as the scale at which turbulence
becomes {\it transAlfvenic}. Thus defined, $L_{MHD}$ can be associated with the scale $l_A$ for 
the transition to the MHD regime of transAlfvenic turbulence that we quantified in this paper. The extension of the FG04 approach to strong subAlvenic turbulence is problematic, however. No quantitative expressions of this $L_{MHD}$ are given in FG04 but the paper contains a footnote "Turbulence can also be injected at smaller velocities on smaller scales, in which case $L_{MHD}$ should be considered an extrapolation beyond the actual outer scale of the cascade." This extrapolation has not been elaborated and it faces conceptual difficulties. Indeed, as we discussed in \S 2, the subAlfvenic turbulence has two regimes,
weak turbulence and strong turbulence. The regime of {\it strong} subAlfvenic turbulence is very
different from the case of transAlfvenic turbulence. In transition to strong subAlfvenic turbulence happens at the scale
$t_{trans}$ and the energy is being injected {\it anisotropically} at this scale, which is in contrast to the isotropic injection
for the transAflvenic turbulence. At scales larger that $l_{trans}$ the turbulence is not any more strong, but follows a
weak turbulence cascade with a very different scaling (see Table 1). Thus there is no physically justified way of defining
$L_{MHD}$ for the subAlfvenic injection of energy in the system. Nevertheless, when expressed in terms of the energy
dissipation\footnote{We find this way of presenting results may sometimes be confusing, as in many cases the turbulent dissipation is not directly measurable in view of
multiple sources of media heating. On the
contrary, the scale of the turbulence $L$ and the magnetic Mach number $M_A$ can be observationally measured as we discuss in this paper.} our results look for strong subAlfvenic turbulence similar to those in FG04, with the difference that the 
weak cascading rather than strong cascading rate enter the formulae. This coincidence stems from the fact that in this
particular regime the damping does depend on the dissipation rate only and not on the injection scale. This makes the case
of strong turbulence special, as in other regimes both the turbulence dissipation rate and the injection scale influence the streaming instability damping.

We have quantified the streaming instability damping for a variety of different regimes of turbulence, including (a) hydro-like superAlfvenic, (b) magnetic superAlfvenic, (c) weak turbulence subAlfvenic, (d) strong turbulence subAlfvenic. The case (b) coincides with the one in FG04 if we identify $L_{MHD}$ there with $l_A$ in this paper.
Damping of streaming in stability in different turbulent regimes are important for different astrophysical environments. For instance, we identified weak turbulence as the major agent for streaming instability damping in subAlfvenic turbulence in the Milky Way halo.  

In terms of astrophysical consequences, we believe that there are no reasons to claim of the catastrophic suppression of the streaming instability by turbulence in galactic environments and therefore do not agree with the conclusion in FG04 related to the crisis in explaining observed degree of isotropy of the Milky Way CRs. Indeed, for the parameters expected for the turbulence in the galactic halo, we found that the turbulent damping should arise
from the interaction of CRs with weak turbulence, rather than with strong turbulence as it is assumed in FG04. This reduces the damping. We provided arguments suggesting that the estimate of the turbulence dissipation rate in FG04 that is based on the cooling of the hot gas is, in fact, an upper limit, which does not constrain the actual turbulence dissipation rate. Therefore we do not believe that the streaming instability must be suppressed by turbulence in the Galactic halo. We also pointed out to the self-regulating nature of the competing non-linear Landau damping of the CR streaming instability. This damping shuts out as soon as freely streaming particles spread into space decreasing the gradients in CR distribution. Thus we conclude that there is no evidence to claim that the streaming instability is suppressed in the Milky Way and therefore it cannot isotropize CRs.

The astrophysical implications of our study is not limited by CR isotropization, however. The expressions that we obtained for the damping of Alfven waves emitted by macroscopic sources describe new ways for launching stellar/galactic winds and heating cosmic plasmas.

\section{Discussion of results}

In this paper we have presented the calculations of Alfven wave damping arising from Alfvenic turbulence. We have dealt both with the case of Alfven waves generated in the local system of reference, as this is the case of Alfvenic waves generated e.g. by streaming instability, and with the case of Alfven waves generated by an external source, e.g. by magnetic perturbations in stellar atmospheres. We have provided the study for a variety of possible astrophysical conditions from superAlfvenic turbulence, i.e. for $M_A>1$ to subAlfvenic turbulence, i.e. for $M_A<1$. We have shown significant changes of wave damping depending on $M_A$ and point out the difference in Alfven wave damping for waves generated in the local system of reference and launched with respect to the mean magnetic field. We have demonstrated that some of the paradoxes noted in the literature disappear when the variations of the turbulence magnetization are taken into account. In particular, we have demonstrated that  the streaming instability can be present in the galactic halo, allowing isotropization of CRs in the Milky Way. The different regimes of damping that we have considered in the paper are applicable to various astrophysical settings and should be accounted for within the detailed modeling. 

Some of our results are presented in a concise form in Table~1. This table describes both the regimes of turbulence and the damping rates for Alfven waves that this turbulence entails. We see, that, compared to the earlier study in FG04, a variety of different scalings are present. The Table also describes the ranges of applicability of different regimes of turbulent damping. Both the damping of waves in the local system of reference, corresponding to the waves generated by streaming instability and damping of waves emitted by external sources parallel to the mean magnetic field are presented. In particular, Table 1 illustrates that the scalings of damping in the two situations and the ranges of the waves for which  damping is applicable are different (see the two
last columns, the first provides the damping of the streaming instability and the range of the CR Larmor radii $r_L$ for which this damping works, the second 
column is for the damping of the waves launched by the external source parallel to the mean magnetic field and the range of the wavelengths for which the damping acts).
Other cases, e.g. Alfven waves emitted at an arbitrary angle, as well as damping of the Alfven waves by outer-scale turbulence are also presented in the current paper.  
We would like to stress the important role of weak turbulence for the suppression of the streaming instability at low $M_A$. While the weak turbulence has a limited inertial range $[LM_A, L]$, it can affect CR streaming for $r_L$ in the range $[LM_A^4, LM_A]$. For instance, for a moderate $M_A=0.1$, the weak turbulence that is present over one decade range of scales can control the propagation of CRs over 3 decades of energy scales. The range of energies of cosmic rays whose streaming is affected by strong subAlfvenic turbulence is significantly reduced. Thus, as we discussed in \S 7.1,  for the Milky Way galactic halo we expect most of the CRs streaming to interact with weak rather than strong turbulence. 

\begin{table*}
\caption{Regimes of MHD turbulence and turbulent Alfven wave damping}
\centering
\begin{tabular}{cccccccc} 
\multicolumn{8}{}{{\bf Table 1}} \\
\multicolumn{8}{}{Non-linear damping of Alfven Waves by MHD Turbulence} \\
\hline
\hline
Type                       & Injection  &  Range                & Spectrum   &                 &  & Instability damping rate & Wave damping rate \\
of MHD turbulence & velocity  & of scales              & E(k)           &                    && and $r_L$ range &  and wavelength range\\
\hline
Weak                      & $V_L<V_A$ & $[l_{trans}, L]$   & $k_{\bot}^{-2}$ &  &  & $\frac{V_A M_A^{8/3}}{r_L^{2/3} L^{1/3}}$, ~~~ $LM_A^4<r_L<LM_A$ & $\frac{V_A M_A^3}{\lambda}$,~~~ $LM_A^3<\lambda< LM_A^2$ \\
\hline
Strong         &                   &                                  &                            &    &               &                     &\\
subAlfvenic& $V_L<V_A$ & $[l_{min}, l_{trans}]$ &  $k_{\bot}^{-5/3}$ &  &  & $\frac{V_A M_A^2}{r_L^{1/2} L^{1/2}}$,~~~ $\frac{l_{min}^{4/3}}{L^{1/3}} <r_L<LM_A^4$&  $\frac{V_A M_A^2}{\lambda^{2/3}L^{1/3}}$,~~~ $l_{min} M_A<\lambda<LM_A^3$ \\
\hline
Hydro-like             &                    &                   &                                      &                                & &                 &  \\
superAlfvenic & $V_L> V_A$ & $[l_A, L]$ &        $k^{-5/3}$        &   &  & $\frac{V_A M_A}{r_L^{2/3} L^{1/3}}$, ~~~ $l_A<r_L<L$ &     $\frac{V_A M_A}{\lambda^{2/3} L^{1/3}}$,~~~  $l_A<\lambda<L$    \\
\hline
Strong           &                      &                   &                                          &      &             &           &     \\
superAlfvenic & $V_L> V_A$ & $[l_{min}, l_A]$ &   $k_{\bot}^{-5/3}$   &  &      & $\frac{V_A M_A^{3/2}}{r_L^{1/2}L^{1/2}}$,~~~  $\frac{l_{min}^{4/3}}{L^{1/3}} M_A<r_L<l_A$ &  $\frac{V_A M_A \sin^{2/3}\theta}{\lambda^{2/3} L^{1/3}}$,~~~ $l_{min}\sin \theta<\lambda<l_A$ \\
\hline
& & & &&&&\\
\multicolumn{8}{l}{\footnotesize{$L$ and $l_{min}$ are the injection and perpendicular dissipation scales, respectively. $M_A\equiv \delta B/B$, $l_{trans}=LM_A^2$ for $M_A<1$  and $l_{A}=LM_A^{-3}$.}}\\
\multicolumn{8}{l}{\footnotesize{for $M_A<1$. For weak Alfvenic turbulence $\ell_{\|}$ does not change. The waves are sent parallel to the mean field, $\theta$ varies as discussed in \S 5.3.}}\\
\end{tabular}
\end{table*}

Our study employs a number of simplifying assumptions the importance of which we would like to discuss. The first of them is that we can consider Alfvenic turbulence separately from the turbulence induced by other modes. This issue has been studied theoretically and quantified numerically (GS95, Lithwick \& Goldreich 2001, Cho \& Lazarian 2002, 2003). The rates of transfer of energy from Alfven to compressible (fast and slow) modes did not exceed the 10\% to 15\% in the study of Cho \& Lazarian (2002). This is a reasonable degree of accuracy for the approximation we employed here.

A more serious point is related to the model of turbulence that is chosen. Our study makes use of the theory of balanced MHD turbulence,\footnote{While there are still debates regarding the nature of this turbulence, we do not believe that there is any evidence in favor of the modifications of GS95 theory that have been suggested so far (Boldyrev 2005, 2006, Beresnyak \& Lazarian 2006, Gogoberidze 2007). These attempts were taken in order to explain low resolution numerical simulations that were getting a power spectrum of Alfvenic turbulence that was systematically more shallow than the $k^{-5/3}$ spectrum suggested by GS95. It was shown, however, in Beresnyak \& Lazarian (2010) that MHD turbulence is less local than hydro and therefore higher resolution numerical simulations are required in order to reveal the actual spectrum of MHD turbulence. Later simulations by Beresnyak (2014) supported this. In view of this we do not find it useful to present our results in terms of modified MHD theories, although it would be very easy to do so.} i.e. when the flow of energy in the opposite direction is the same, while localized astrophysical sources and sinks of turbulent energy may make Alfvenic turbulence imbalanced, i.e. with the flow of energy in one direction exceeding the flow in the opposite direction. Solar wind up to 1AU presents an example of such imbalanced turbulence. A few theories have been suggested to account for imbalance (e.g. Lithwick \& Goldreich 2007, 
Beresnyak \& Lazarain 2008b, Chandran 2008, Perez \& Boldyrev 2009). Among these theories, the one by Beresnyak \& Lazarian (2008b) was shown to correspond to numerical simulations in Beresnyak \& Lazarian (2009). For small imbalances, this theory smoothly transfers to GS95, while large imbalances are difficult to create in astrophysical media because of reflection of Alfvenic perturbations in realistically compressible and inhomogeneous media. Therefore we believe that our present study can provide a reasonable estimate for such situations.\footnote{When the imbalance of turbulence is important, our treatment can be generalized using the anisotropies of the component of Alfvenic turbulence propagating in the opposite direction to wave propagation. For instance, Beresnyak \& Lazarian (2008b) predicted different anisotropies for stronger and weaker oppositely propagating 
components.}

Our treatment was presented for a single scale of energy injection. In real astrophysical situations small scale energy injection
takes place along with the large scale energy cascade. The local energy injection may dominate the dynamics of small scale eddies. For instance, locally, turbulence in the precursor of supernovae shocks definitely dominates the turbulence of the large scale Galactic cascade. Our treatment can be generalized for such situations.

This study has important astrophysical implications. Naturally, astrophysical fluids exhibit a variety of turbulent regimes. The damping of the waves also depends on how the Alfven waves are launched. We have quantified the whole variety of the regimes of Alfven wave damping by strong and weak subAlfvenic turbulence, turbulence at the injection scale and at the dissipation scale. We identified the difference in damping for the waves emitted by streaming particles and macroscopic astrophysical sources. The latter are essential e.g. for launching stellar and galactic winds, or for the heating of intracluster media. Our exploration of the damping of the streaming by superAlfvenic turbulence is the closest to that described in the earlier studies.  For a number of implications we have just sketched the possible physics and do not get into the quantitative details. This is natural,
as the Alfven wave damping in turbulent media is widely spread in astrophysical settings and this paper is focused on quantifying different regimes of the process rather than
its numerous astrophysical consequences. 

We would like to stress that our paper is focused on turbulent damping of Alfven waves. We deal with the non-linear Landau damping (see Kulsrud 2005) only to the extent that is required for the purpose of the comparison of the importance of two mechanisms for
the damping of the streaming instability. We argue that the non-linear Landau  damping
acts in self-regulating fashion and therefore it can be in many instances subdominant compared to the damping by turbulence. Indeed, if the streaming instability is suppressed by the aforementioned mechanism, it allows the spread of the CRs, increasing
the scale hight of their distribution. This, in its turn, suppresses the non-linear Landau damping. On the basis of this reasoning, we believe that in most astrophysical
situations turbulent damping is more important that the non-linear Landau damping of Alfven waves. 

\section{Summary}

Our study of Alfven wave damping in MHD turbulence revealed a variety of damping regimes with important astrophysical consequences. We express our results through the magnetization of the media, which is given by the Alfven Mach number $M_A$ and the turbulence injection scale $L$. Both quantities can be obtained through observations. We quantified the wave damping for different regimes of subAlfvenic, superAlfvenic turbulence as well as for damping of Alfven waves by the turbulence at the injection scale. In every case we obtained the range of wavelengths for which the damping  in the particular regime is applicable. This work opens ways for studies of the consequences of Alfven wave damping in various astrophysical settings. Those include launching of stellar and galactic winds, heating of the media, control of the CR streaming instability, etc. Our results can be briefly summarized as follows:
\begin{itemize}
\item The damping is different if Alfven waves are generated in the local system of reference and in the global system of reference when Alfven waves are launched with respect to the mean field.  The former case takes place when e.g. when particles subject to the streaming instability generate waves with respect to the magnetic field that they interact with, while the latter case takes place e.g. when Alfven waves are injected into turbulent media by a external macroscopic source.  The structure of the Alfven wavefronts and their interaction with turbulence in two cases is different and this entails the difference in turbulent damping.
\item For Alfven waves launched by streaming instability, their damping in subAlfvenic turbulence is different for weak and strong regimes of turbulence. In both cases, the damping is significantly slower compared to the case of the superAlfvenic turbulence. The weak turbulence, while being present over a limited range of scales, can affects CR streaming over a significant range of energies. On the contrary, the range of the damping by strong subAlfvenic turbulence is significantly reduced. This results, for instance, that the CR streaming in low $M_A$ environments of galactic halos is mostly affected by weak turbulence.
\item The damping of Alfven waves launched with respect to the mean magnetic field depends on the angle between the mean field and the direction of the wave propagation. In the limiting case of Alfven waves propagating along the mean magnetic field, the damping is different compared to that present for the waves send along the local direction of magnetic field by the streaming instability. 
\item The study suggests many important astrophysical consequences, many of which are still to be elaborated. For instance, the efficient damping of Alfven waves generated by astrophysical sources, e.g. stars and galaxies, provides heating of the media and launching of winds as Alfven waves deposit both their energy and momentum into the ambient turbulent magnetized plasmas. At the same time, we do not expect the damping of streaming instability by turbulence in the Milky Way galactic halo to be to be strong enough to affect the observed level of galactic CR isotropy.  
\end{itemize}

\acknowledgments
I acknowledge the NSF grant AST 1212096, NASA grant NNX14AJ53G and the NSF Center for Magnetic Self Organization (CMSO) as well as  a distinguished visitor PVE/CAPES appointment at the Physics Graduate Program of the Federal University of Rio Grande do Norte, the INCT INEspao and Physics Graduate Program/UFRN. Productive and stimulating discussions with Jungyeon Cho and Gianfranco Brunetti as well as helpful comments of the anonymous referee are acknowledged.

\clearpage

\end{document}